\documentclass[twocolumn, preprintnumbers, prl, floatfix, aps]{revtex4-1}
\usepackage{amssymb,amsmath,amsfonts}
\usepackage{graphicx,epsfig} 
\usepackage{bm} 
\usepackage{color}

\definecolor{Myorange}{cmyk}{0,0.42,1,0}

\definecolor{amaranth}{rgb}{0.9, 0.17, 0.31}

\usepackage{soul}
\begin{document}
\title{Ordinal spectrum: a frequency domain characterization of complex time series}
\author{Mario Chavez}
\email[Electronic adress: ]{neurodynamicslab@gmail.com}
\affiliation{CNRS UMR-7225, H\^{o}pital de la Piti\'e-Salp\^{e}tri\`{e}re. Paris, France}%
\author{Johann H. Mart\'inez}
\affiliation{Grupo Interdisciplinar de Sistemas Complejos (GISC). Madrid, Spain}%
\date{\today}
\begin{abstract}
Although classical spectral analysis is a natural approach to characterise linear systems, it cannot describe a chaotic dynamics. Here, we propose the {\em ordinal spectrum}, a method based on a spectral transformation of symbolic sequences, to characterise the complexity of a time series. In contrasts with other nonlinear mapping functions (e.g. the state-space reconstruction) the proposed representation is a natural approach to distinguish, in a frequency domain, a chaotic behavior. We test the method in different synthetic and real-world data. Our results suggest that the proposed approach may provide new insights into the non-linear oscillations observed in different real data. \end{abstract}
\maketitle %

Observed time series from a large number of physical processes generally display erratic temporal behavior.  In the last decades, various measures of complexity have been proposed to characterize these data and distinguish regular (e.g., periodic), chaotic, and random dynamics~\cite{abarbanel1993}. The spectral representation based on the Fourier transform could be a natural approach to identify the rich oscillatory dynamics in terms of frequency, magnitude and phase. Indeed, changes in power spectra of some dynamical systems as they bifurcate to chaos have been described~\cite{crutchfield1980, farmer1981, dumont1988}. Although several measures were proposed to characterize these spectral changes~\cite{crutchfield1980, lafon1983, koussoulas2001, wiebe2012}, a close relation between these measures and the attractor topology is unclear and it is not supported by numerical simulations~\cite{crutchfield1980, lafon1983}.   

It is generally accepted that time series observed from chaotic systems exhibit some characteristic signatures such as fractal geometry and broadband frequency content.  Indeed, time series of chaotic trajectories often display an exponential decay in their power spectrum at high frequencies, different from the classical power-law of stochastic colored noises~\cite{sigeti1987, sigeti1995, wiebe2012}.  However, some studies have proved that the spectra of colored noises cannot be related to the dynamical route to chaos~\cite{white1996a, white1996b}. In fact, the power spectrum estimated from chaotic sequences can be replicated by a monotonic nonlinear transformation of linearly filtered noise~\cite{theiler1992, schreiber2000, small2003}. All this evidence suggests that classical spectral analysis cannot provide adequate information for identifying chaotic systems. 

Although classical spectral analysis is adequate for characterizing linear systems, power spectra cannot reflect the non-linear interactions between the Fourier components of a chaotic motion~\cite{farmer1981, subbarao1992, chandran1993}. Bispectral techniques has been used to investigate such nonlinear interactions~\cite{subbarao1992, chandran1993}, but some works have proved that higher-order statistics are required for a better characterization of a chaotic dynamics~\cite{pezeshki1990, chandran1993}. Recently, other nonlinear spectral methods have been proposed to distinguish deterministic from stochastic dynamics in finite time series. The so-called symbol spectrum test developed in Refs.~\cite{kulp2011, kulp2014} does not take into account the temporal dynamics of the symbols, but the variability of their distribution in the symbolic sequence. The spectrum proposed in Ref.~\cite{sun2007} characterises, in the frequency domain, the recurrence of a reconstructed trajectory in the phase space. 

Based on the concept of state-space reconstruction, some measures such as entropies, Lyapunov exponents and fractal dimensions were shown to be effective to characterize and reconstruct the equations of motion when the data model has a deterministic dynamics~\cite{abarbanel1993, sprott2003}.  Alternative nonlinear mapping functions were proposed to better capture the disorder degree of a time series through symbolization procedures~\cite{mischaikow1999, demicco2008}. The method known as ordinal patterns (OP) is a transformation based on order relations among values of the data, and it provides a robust estimation of the probability distribution function associated with the time series~\cite{bandt2002}. This representation of the ordinal structure has provided a robust tool to discriminate, in the time domain, different dynamical regimes in time series~\cite{zanin2012, amigo2015}. Coarse-graining approaches~\cite{costa2002, humeau2015}, or the use of ordinal structures for different time delays~\cite{zunino2012} have been proposed to characterize complexity at different temporal scales. 

In this Letter we present the {\em ordinal spectrum} of time series. The method is based on the spectral transformation of a symbolic representation of data. The method is based on the ordinal patterns and it is therefore fully data-driven. In contrast with other nonlinear approaches, the spectral analysis proposed here provides a characterization of the data's complexity in a frequency domain. We assess the reliability of our method in distinguishing periodic or random time series from chaotic data. We evaluate the proposed method in different synthetic and real, linear, non-linear, stochastic and deterministic time series. Results depict a robust approach to identify a chaotic dynamics in data.

The main steps to estimate the ordinal spectrum from a time series $X_t$ are the following:

\emph{i) Ordinal pattern representation of data}. Symbolisation procedures map a time series $X_t$ onto a discretized symbols sequence by extracting its amplitudes' information~\cite{amigoBook}. Among several symbolisation proposals \cite{daw2003}, we considered here the dynamical transformation of OP~\cite{bandt2002}. This method maps a time series $X_t$ with $t = 1,\ldots,T$ to a finite number of patterns that encode the relative amplitudes observed in the $D$-dimensional vectors $\mathbf{X}_t=\{X_t, X_{t+\tau}, \ldots, X_{t+(D-1)\tau}\}$. The elements of the vector $\mathbf{X}_t$ are mapped uniquely onto the permutation $\mathbf{\pi}  =(\pi_0,\pi_1,\ldots,\pi_{D-1})$ of $(0,1,\ldots,D-1)$ that fulfills $X_{t+\pi_0\tau} \leqslant X_{t+\pi_1\tau}, \leqslant \ldots \leqslant X_{t+\pi_{D-1}\tau}$. 

The set of all possible ordinal patterns derived from a time series, that represents the whole embedding state space, is noted as $Z$, whose cardinality is $D!$ at most. The whole sequence of OP extracted from $X_t$ is known as the symbolic representation $S_t$ of the time series. The higher the order is, the more information is captured from the time series. To sample the empirical distribution of ordinal patterns densely enough for a reliable estimation of its probability distribution we follow the condition~\cite{amigoBook} $T\geqslant (D+1)!$. The OP symbolisation has some practical advantages~\cite{zanin2012, amigo2015}: \textit{a)} it is computationally efficient, \textit{b)} it is fully data-driven with no further assumptions about the data range to find appropriate partitions, \textit{c)} it is invariant to any monotonous transformations and, \textit{d)} a small $D$ is generally useful in descriptive data analysis~\cite{bandt2002, amigoBook}. Furthermore, this representation is known to be relatively robust against noise, and useful for time series with weak stationarity~\cite{amigoBook, zanin2012, rosso2007, keller2014}. 

Contrary to phase state reconstruction, in ordinal time-series analysis, the criteria to select an embedding dimension $D$ are computational cost and statistical significance in view of the amount of data available. The selection of time delay embedding $\tau$ may, however, influence
the analysis of correlated data.  Here, to minimize the effects of this correlation we select the delay that corresponds to either the first minimum or the zero crossing of the autocorrelation function $\rho$ of the original time series $X_t$ (the folding time $\rho=1/e$ is used in case of a monotonic function $\rho$).

\emph{ii) Capturing information dynamics from the symbolic sequence}. To characterize the time evolution of the obtained sequence $S_t$ we consider it to be a homogeneous ergodic Markov chain with the finite state space $Z = {z_1,\ldots, z_n}$ (all the possible distinct permutations). Let $P^m = \{p^m_{ij}\}$ be the transition matrix that describes the probability of leaving the symbol $z_i$ and entering the symbol $z_j$ at a distance $m$, i.e. $p^m_{ij} = p(S_{t+m}=z_j | S_{t}=z_i)$.  If the chain is stationary we have $p(S_t=z_i)= p^*_i$, where $p^*_i$ is the invariant or stationary distribution that satisfies $p^*_j = \sum_i p^*_i p_{ij}$ 

\emph{iii) Characterization of symbolic dynamics at different time lags}. For a first-order Markov chain, any symbol in sequence $S_t$ is independent of all the previous observations. However, by construction, each ordinal pattern in $S_t$ depends on its predecessors, inducing thus a non-zero correlation between symbols $S_t$ and $S_{t+m}$ for $m>1$. The correlation between $S_t$ and $S_{t+m}$, is termed the autocovariance at lag $m$ and, for a Markov chain that converges to a unique stationary distribution, it is expected to decrease as the lag is increased~\cite{basawa1972, fuh1992}. 

We notice that, in contrast with numeric signals, symbolic sets have no mathematical structure and algebraic operations are usually meaningful. To solve this, different rules have been proposed for mapping a symbolic sequence into a numerical domain~\cite{voss1992, wang2009}. The numerical algorithm used in the ordinal pattern transformation also yields an enumeration of permutations, such that each unique ordinal pattern can be associated to a non-negative integer, $z_i=i$ with $i \in \{1, \ldots, D! \}$~\cite{keller2007, berger2019}. Interestingly, this enumeration procedure yields a natural order of the symbols such that the pattern with permutation representation $(1,\ldots, D)$ and that with representation $(D,\ldots, 1)$ are at the maximal possible distance since they represent completely opposite monotonic behaviors~\cite{keller2007}.  

On the basis of this representation with ordered symbols or patterns, a rank variance and rank autocovariance of the Markov chain can be obtained as follows~\cite{basawa1972, fuh1992}:
\begin{equation} \label{varaincecovarianceMarkov}
\begin{split}
\mathrm{Var}(S) &=\sum_i i^2 p(S_t=z_i) - \mathrm{E}\{S_t\}^2 \\
\mathrm{Cov}(m) & = \mathrm{Cov}(S_t,S_{t+m}) \\ 
			& = \sum_i \sum_j i j p(S_t=z_i, S_{t+m}=z_j) \\
			& \phantom{\mathrm{E}\{S_t\}^2 } - \mathrm{E}\{ S_t\}^2,
\end{split}
\end{equation}
where $\mathrm{E}\{S_t\} = \sum_i i p(S_t=z_i)$. We can also write $p(S_t=z_i, S_{t+m}=z_j) = p^*_i p^m_{ij}$, wich yields:
\begin{equation} \label{covarianceMarkov}
\mathrm{Cov}(m) = \sum_i \sum_j i j p^*_i p^m_{ij} - \left( \sum_i i p^*_i \right)^2
\end{equation}

\emph{iv) Estimation of the ordinal spectrum}. To explore the spectral properties of the ordinal patterns sequence, the \emph{ordinal spectrum} (OS) can be obtained from the spectral representation of the autocovariance function defined above:
\begin{equation}
\label{OS}
\mathrm{OS}(f) =\sum_{m=-(N-1)}^{N-1}\mathrm{Cov}(m) \exp(-i 2\pi f m)
\end{equation}

Periodicity in time series yields a periodic structure in the symbolic sequences, and will be reflected in an ordinal spectrum with clear peaks. Similarly, as in their counterpart in classical signal processing, random symbolic sequence are decorrelated with a flat spectrum. It is well known that the structure of symbolic sequences depends on the temporal correlations of original time series~\cite{carpi2010}. Consequently, their symbolic autocovariance and ordinal spectrum are expected to depend on the degree of such correlations. Although different estimators can be used to obtain the spectrum $\mathrm{OS}(f)$ from the covariance function $\mathrm{Cov}(S_t,S_{t+m})$, here we used the autoregressive spectral estimate as it offers better spectral resolution and smaller variance than does Fourier-based estimators~\cite{therrien1992, ARcoefficients}. 
 
\begin{figure}[!htp]
   \centering
   \resizebox{\columnwidth}{!}{\includegraphics{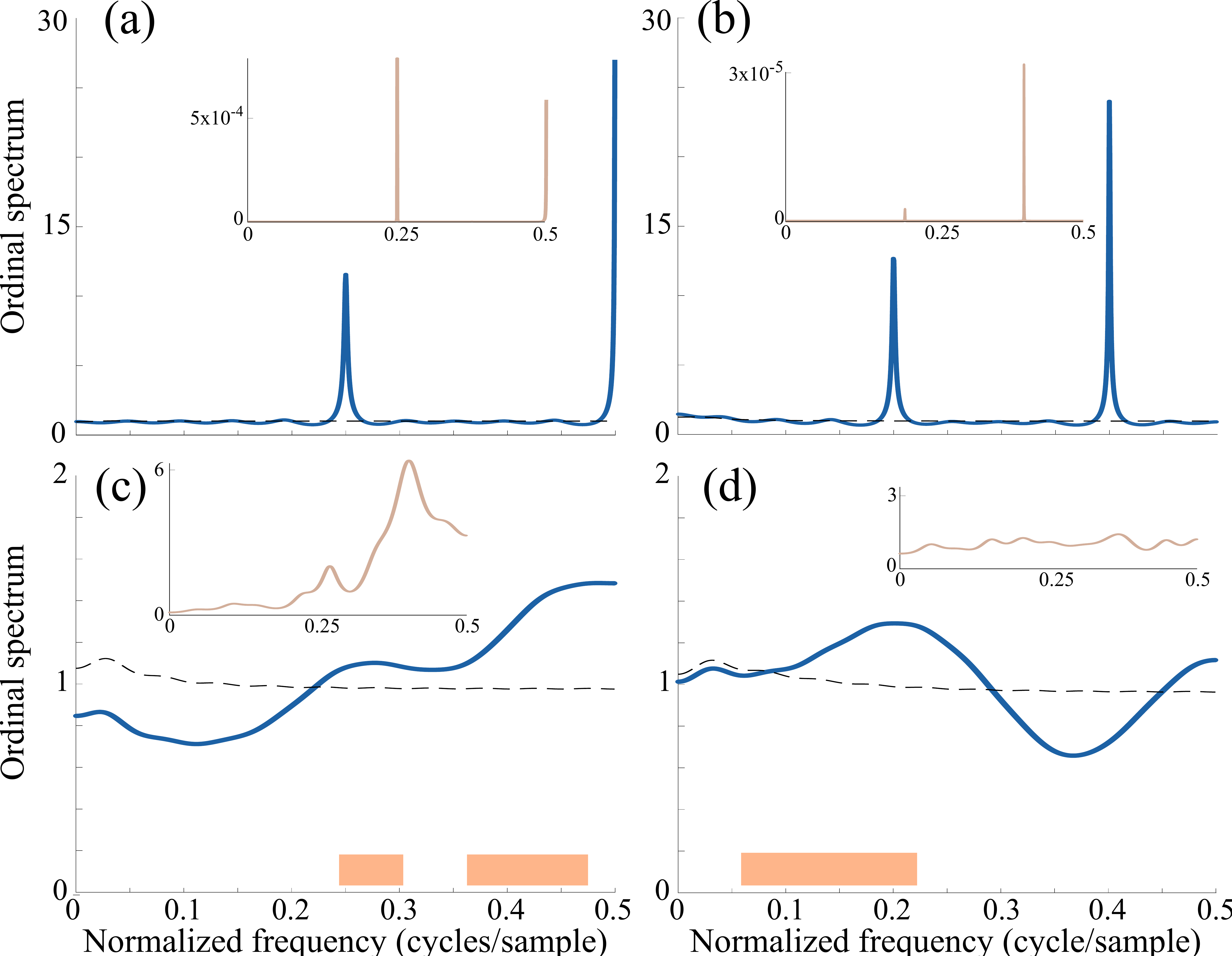}}
   \caption{Ordinal spectra against normalised frequency (blue curves) for data generated from the logistics map (embedding parameters $D=3$ and $\tau = 1$): with periodic behaviors (a) $r=3.55$; (b) $r = 3.739$; and chaotic dynamics (c) $r=3.8$; and (d) $r=4$. Orange boxes in main subplots indicate the frequency regions where values of ordinal spectra are statistically different from those obtained from surrogate data. Slashed curves only indicate the average spectra from randomized symbolic sequences, not from IAAFT surrogates. Insets depict the power spectra from original time series in normalised frequencies.}  
   \label{OSlogisticMaps}
 \end{figure}

\emph{v) Detection of relevant scales in the ordinal spectrum}. Peaks in the ordinal spectrum could simply result from large autocorrelation values at different time lags in the original time series $X_t$. To rule out this possibility, we compare the ordinal spectrum with those obtained from an ensemble $\{X_t^s\}$ of surrogate time series that replicate the linear autocorrelation and amplitudes distribution of the original time series. Here, we use the so-called Iterative Amplitude Adjusted Fourier Transform (IAAFT)~\cite{schreiber2000, small2003} that preserves autocorrelation function and amplitude distribution of original data, while all other higher-order statistics are destroyed. For each $X_t^s$, we repeat the above steps \textit{i)-iv)} to compute a set of $\{\mathrm{OS}^s(f)\}$ spectra. If any value in the ordinal spectrum $\mathrm{OS}(f)$ is statistically distant from the distribution of $\{\mathrm{OS}^s(f)\}$ we can reject the null hypothesis of a linear stochastic time series. In this study, all significance tests are set at $p < 0.05$, Bonferroni-corrected for multiple comparisons (over frequencies $f$). 

\begin{figure}[!htbp]
   \centering
   \resizebox{\columnwidth}{!}{\includegraphics{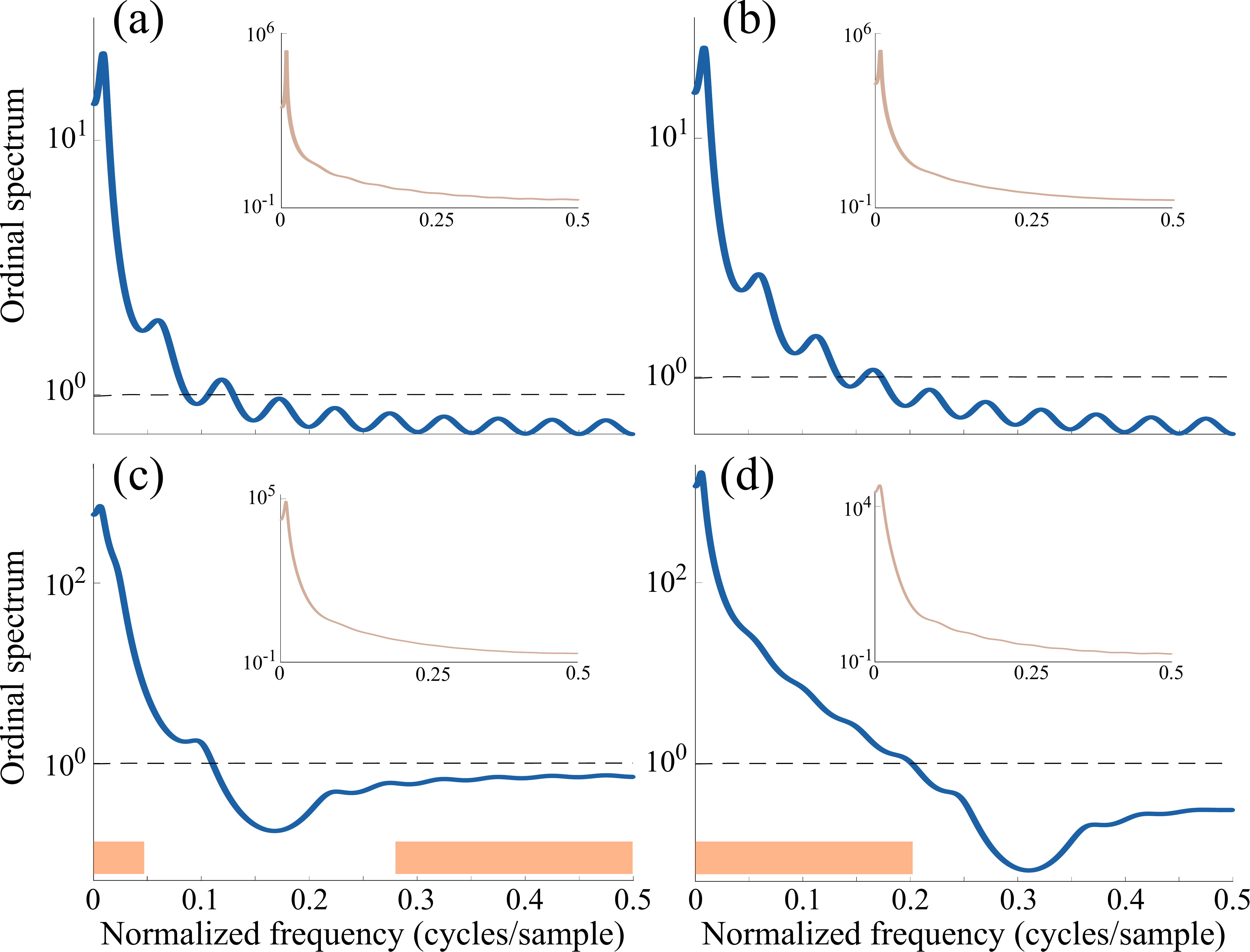}}
   \caption{Ordinal spectra against normalised frequency for data generated from the $x$-component of the R\"ossler system ($D=4$ and $\tau = 30$ samples) with: periodic behaviors (a) $a=0.30$; (b) $a=0.41$; and chaotic dynamics (c) $a=0.42$; and (d) $a=0.54$. Same stipulations as in the caption of Fig.\ref{OSlogisticMaps}}  
   \label{OSrossler}
 \end{figure}

To demonstrate our method, we apply it first to synthetic data generated by a logistic map, defined by the iterative equation $x_{n+1} = rx_n(1-x_n)$ where $r$ is the bifurcation parameter. This nonlinear map has several transitions in the dynamics occurring during $r\in \left[3, 4\right)$, with several period-doubling cascades before the onset of chaos at $r\approx 3.56995$~\cite{may76}. Beyond this value, several isolated ranges of $r$ display non-chaotic behavior~\cite{may76}. For this model, the length of each time series is set to $T = 2000$, after discarding the first 1000 points to avoid possible transients. 
 
Main plots in Figs.\ref{OSlogisticMaps}(a)-(b) show that, despite the large peaks observed in the ordinal spectrum, the dynamical properties of the periodic process are not statistically different from those replicated by the surrogate data and thus, the null hypothesis cannot be rejected at any frequency. As expected for chaotic sequences, results in Figs.\ref{OSlogisticMaps}(c)-(d) indicate that the ordinal spectrum captures in different frequency ranges a dynamical complexity different from those of surrogates. 

We also test our method on data generated by the R\"ossler system whose equations are given by $\left[ \dot{x}=-y-z, \quad \dot{y}=x+ay, \quad \dot{z}=2+z(x-4) \right]$, with the control parameter $a \in [0.25, 0.55]$. Similar to the logistic map, R\"ossler system has several periodic transitions before the onset of chaos at $a \approx 0.385$. For larger values of $a$, some periodic windows can be still found. The length of each time series is set to $T = 10^4$, and a transient cut-off of $1000$ samples. 

Different plots in Fig.\ref{OSrossler} show that the proposed method accurately distinguish chaotic fluctuations from a periodic dynamics (event for the dynamics observed in the pocket of periodicity at $r=0.41$). Whereas original time series display a similar power law decay in their spectra (identical, by construction, to those obtained from surrogate data), the ordinal spectrum capture a dynamical complexity that can not be replicated by the surrogates.

To further evaluate our method, we consider short sample sizes. Numerical simulations show that the ordinal spectrum test can correctly detect chaos in the logistic map when the data length is, at least, $T = 500$ samples. For the chaotic R\"ossler system, the method requires the time series to be larger than ten times the fundamental period of the oscillation. For shorter sample sizes, the method cannot reject the null hypothesis of a linearly filtered noise.

In addition to testing across deterministic time series (chaotic or not), we evaluated the method on non-chaotic stochastic data. Here we firstly evaluated a Gaussian noise with distribution $\mathcal{N}(0,1)$ and a stochastic process with a power law spectrum $S(f) \propto f^{-\alpha}$ where $\alpha =1$. For each stochastic process, the ordinal spectrum is calculated with 2000 samples.

\begin{figure}[!ht]
   \centering
   \resizebox{\columnwidth}{!}{\includegraphics{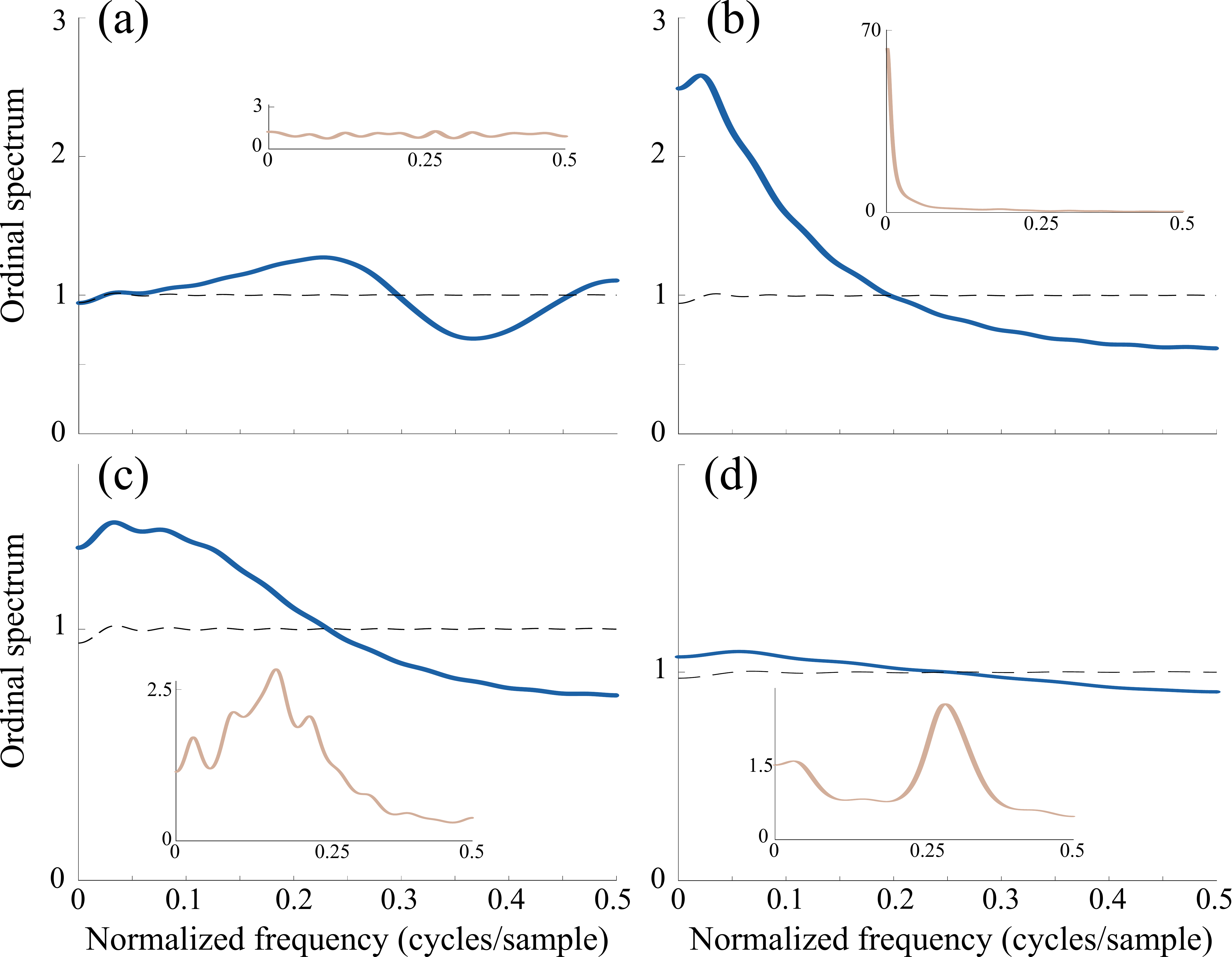}}
   \caption{Ordinal spectra for stochastic data generated from: (a) a Gaussian distribution ($D=4$ and $\tau = 1$ samples); (b) Power law noise ($D=4$ and $\tau = 4$ samples);  (c) nonlinear system driven by non-Gaussian noise ($D=4$ and $\tau = 2$ samples); and (d) non-monotonic nonlinear transformation of a linearly filtered noise ($D=4$ and $\tau = 3$ samples). Same stipulations as in the caption of Fig.\ref{OSlogisticMaps}}
\label{stochasticSystems}
 \end{figure}

Results in Figs.~\ref{stochasticSystems}(a)-(b) suggest that, although the ordinal spectra are affected by the temporal correlations of the original time series, they cannot be distinguished from those produced by the surrogate data. As expected, the corresponding statistical tests suggest that the null hypothesis of a stochastic process cannot be rejected in both cases (white decorrelated and correlated noises).

We then evaluated the method on a nonlinear system driven by a non-Gaussian noise. We considered the system given by $x_t = 0.5x_{t-1} - 0.3x_{t-2} + 0.1y_{t-2} + 0.1x_{t-2}^2 + 0.4y_{t-1}^2 +  0.0025\eta'_t$ and $y_t=\sin(4\pi t)+\sin(6\pi t) + 0.0025\eta''_t$, where noises $\{\eta'_t$,$\eta''_t\}$ are iid drawn from the Laplacian distribution $p(\eta) = \frac{1}{4b}\exp \left(\frac{-|\eta-\mu|}{b}\right)$, with $\mu=0$ and $b=1$. To evaluate the performance of our method under the null hypothesis of nonlinearly transformed stochastic processes, we applied a static non-monotonic nonlinear transformation, $x_{t} = \tanh^2{(y_{t})}$, to the linear non-Gaussian process given by $y_t = 0.4y_{t-3}-0.3y_{t-2}0.2y_{t-1} + e_t$, where $e_t$ is obtained  by squaring a uniform noise with amplitude distribution between $-0.5$ and $0.5$~\cite{diks1995}. For these two models, the data length is also set to $T = 2000$, after discarding the first 1000 points.

A clear distinction between chaos and stochastic behavior can be difficult for data generated by nonlinear systems driven by non-Gaussian noises~\cite{schreiber2000, small2003}. Similarly, it is well known that nonlinear transformations may introduce sufficient phase correlations in linearly filtered noises making difficult to identify the stochastic behavior. Results depicted in Figs. ~\ref{stochasticSystems}(c)-(d) indicate that the ordinal spectrum, in combination with the IAAFT algorithm, correctly diagnoses the nonlinear and non Gaussian models as stochastic process, including the static nonlinear non-monotonic transformation of a non Gaussian random process.

Results clearly indicate that: \textit{i)} large peaks in the nonlinear spectrum cannot be considered as a proof of a chaotic dynamics; and \textit{ii)} whatever the underlying dynamic is (periodic or chaotic), a random shuffling of the constructed symbolic sequence yields a flat spectrum. A statistical test based on randomly shuffled sequences is therefore unable to identify chaotic dynamics. 

Finally we demonstrate the potentials of our method on real data of different nature: epidemiology (measles and cholera time series), astrophysics (the sunspots number series) and neuroscience (electroencephalographic data from an epileptic patient). As data have different length we apply the ordinal symbolic transformation in different dimensions, following the condition~\cite{amigo2015} $T\geqslant (D+1)!$
 
 \begin{figure*}[!htbp]
   \centering
   \resizebox{\textwidth}{!}{\includegraphics{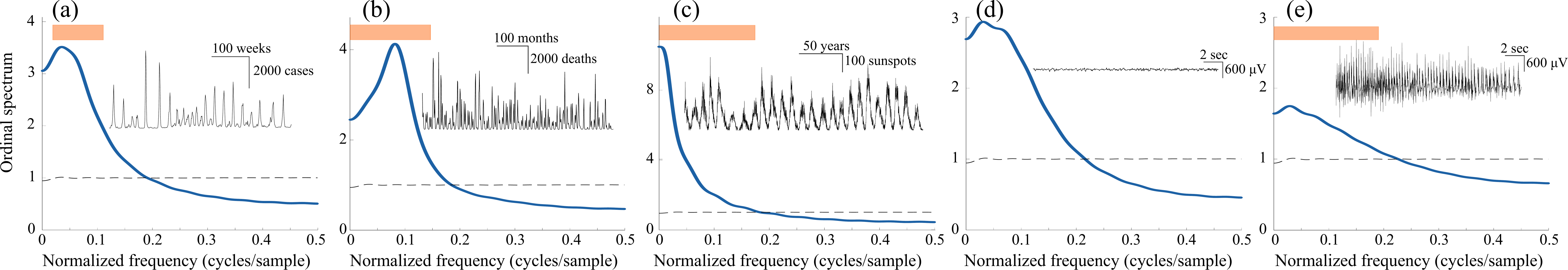}}
   \caption{Ordinal spectra (blue curves) of different real data. Insets show the original time series, its temporal and amplitude scales. (a) Measles cases ($D=3$ and $\tau = 4$ samples); (b) cholera data ($D=4$ and $\tau = 2$ samples);  (c) Sunspots data ($D=4$ and $\tau = 30$ samples);  (d) EEG before and (e) during an epileptic seizure ($D=4$ and $\tau = 5$ samples). Slashed curves indicate the average spectra from randomly shuffled symbolic sequences. Orange boxes in main subplots denote the frequency regions where values of ordinal spectra are statistically different from those obtained from surrogate data.}  
   \label{realData}
 \end{figure*}

The pattern of measles epidemics in developed countries is among the best documented population cycles in ecology. Different studies have proposed evidence for low dimensional dynamics in different epidemiological time series~\cite{olsen1988, bolker1993}. The inset in Fig.~\ref{realData}(a) shows the monthly cases of measles in Copenhague, Denmark, between 1927 and 1968~\cite{olsen1988}. For the time series of measles considered here, our method clearly rejects the null hypothesis of stochastic dynamics. In full agreement with previous works, our results indicate that such measles' dynamics cannot be analyzed by conventional linear models and a low dimensional complexity underlies the observed dynamics.
 
The interannual disease cycles observed in many infectious diseases result from the interplay between extrinsic and intrinsic factors~\cite{koelle2005}. These interactions with the disease dynamics may produce oscillations of complex patterns, including chaos~\cite{schwartz1992}. 
The inset in Fig.~\ref{realData}(b) depicts the monthly deaths from cholera in Dacca, East Bengal between 1891 and 1940~\cite{king2008, choleraData}.
Our analysis, based upon a spectrum capturing the intrinsic nonlinear dynamics of the system, suggests a low dimensional dynamics in the time series. This is in agreement with previous mathematical models of seasonally driven epidemics~\cite{schwartz1992}.

Solar activity is driven by the emergence of magnetic flux through the photosphere forming active regions which include sunspots. While the most characteristic feature of solar activity is its modulated 11-year cycle~\cite{letellier2006}, some studies have suggested that the irregular behavior of the activity reflects the presence of low-dimensional chaotic dynamics~\cite{letellier2006}.  The inset in Fig.~\ref{realData}(c) shows the monthly mean total sunspot number (the arithmetic mean of the daily total sunspot number over all days of each calendar month) between 1749 and 2020~\cite{sunspotsData}. Our method rejects the hypothesis of a stochastic process, which agrees with previous findings suggesting that sunspots fluctuations can be explained by a nonlinear (chaotic) dynamics~\cite{letellier2006}.
 
As many other time series in biology and medicine, electroencephalographic (EEG) signals display strong nonlinearities during different cognitive or pathological states~\cite{stam1998}. In epilepsy, dynamical properties of EEG signals can be a used as a marker of the epileptogenic zone~\cite{andrzejak2001}. Here, we applied our method to a scalp EEG recordings from a subject with intractable epileptic seizures. Data were recorded at 102.4 Hz with a scalp right central (C4) electrode (linked earlobes reference)~\cite{quiroga1997, eegData}. Time series ploted in Figs.~\ref{realData}(d)-(e) correspond to data from interictal and ictal (seizure) period, respectively. These results confirm previous findings suggesting that interictal EEG dynamics can be associate to a stochastic process, whereas a low dimensional dynamics characterizes epileptic seizures~\cite{andrzejak2001}. 
    
To conclude, this study proposes a nonlinear spectrum for characterising complexity in the frequency domain. Our approach is able to distinguish chaotic fluctuations from stochastic dynamics in finite time series. Based on the ordinal patterns analysis, the proposed method compares the spectral information of the symbolic representation of $X_t$ and the counterpart of linearly filtered stochastic process (surrogate data). Our simulations suggest that it accounts for static nonlinear transformations of linear data and accurately provides the expected results, even under the null hypothesis of correlated noises or nonlinearly transformed stochastic processes. 

In this work, we show that the detection of chaotic oscillations in time series can be successfully addressed using a spectral analysis of ordinal symbolic representation.  Our findings depict a robust approach to identify a chaotic dynamics in a time series. The main advantage of our proposal relies on its simplicity, reliability, and computational efficiency. The method is fully data-driven and it does not require  a priori knowledge about the data sequence for its symbolic representation, which is very useful in real-world data analysis. Although it is based on the ordinal patterns representation, the method can be straightforwardly applied to other symbolic representations (based on ordered symbols or categories). Similarly, different spectral representations can be obtained from symbolic data.  

Our results confirm chaotic dynamics in sunspots time series and suggest this property as a common signature in epidemiological data.  Results also indicate that the EEG recordings are characterized by a complex dynamics during epileptic seizures, whereas interictal activity can be explained by a stochastic process. These results suggest that nonlinear spectral methods may provide a more complete characterization of chaotic sequences, which could help to unfold their underlying dynamics and provide a better landscape of the observed system. A spectral description of complex time series might provide, more in general, meaningful insights into the complex oscillations observed in other data such as biomedical, financial or climate time series. 
\\
\begin{acknowledgments}
J.M. thanks the MinCiencias-Colombia postdoctoral research program No 811.
\end{acknowledgments}

\end{document}